\documentclass[prx,aps,twocolumn, amsmath,amssymb,superscriptaddress, notitlepage]{revtex4-1}
\usepackage[dvipsnames]{xcolor}
\usepackage{amssymb}
\usepackage{esint}

\usepackage{amstext}
\usepackage{graphicx,epstopdf}
\usepackage{hyperref}
\usepackage{color}
\usepackage{amsmath}
\usepackage{color}
\usepackage{enumitem}
\usepackage{soul}
\usepackage{makecell}
\usepackage{multirow}
\usepackage{booktabs}
\usepackage{bm}

\makeatletter

\begin{document}

\title{Single trajectory characterization via machine learning}


	\author{Gorka Mu\~noz-Gil}
	\affiliation{ICFO -- Institut de Ci\`encies Fot\`oniques, The Barcelona Institute of Science and Technology, 08860 Castelldefels (Barcelona), Spain}
	\author{Miguel Angel Garcia-March}
	\affiliation{Instituto Universitario de Matem\'atica Pura y Aplicada, Universitat Polit\`ecnica de Val\`encia, E-46022 Val\`encia, Spain}
	\author{Carlo Manzo}
	\affiliation{Facultat de Ci\`encies i Tecnologia, Universitat de Vic -- Universitat Central de Catalunya (UVic-UCC), C. de la Laura,13, 08500 Vic, Spain }
	\author{Jos\'e D. Mart\'in-Guerrero}
	\affiliation{Universitat de Val\`encia -- IDAL, Department of Electronic Engineering, ETSE-UV, Avgda. Universitat, s/n 46100 Burjassot (Valencia), Spain }
	\author{Maciej Lewenstein}
	\affiliation{ICFO -- Institut de Ci\`encies Fot\`oniques, The Barcelona Institute of Science and Technology, 08860 Castelldefels (Barcelona), Spain}
	\affiliation{ICREA, Lluis Companys 23, E-08010 Barcelona, Spain}

\begin{abstract}
	In order to study transport in complex environments, it is extremely important to determine the physical mechanism underlying diffusion and precisely characterize its nature and parameters. Often, this task is strongly impacted by data consisting of trajectories with short length (either due to brief recordings or previous trajectory segmentation) and limited localization precision. In this paper, we propose a machine learning method based on a random forest architecture, which is able to associate single trajectories to the underlying diffusion mechanism with high accuracy. In addition, the algorithm is able to determine the anomalous exponent with a small error, thus inherently providing a classification of the motion as normal or anomalous (sub- or super-diffusion). The method provides highly accurate outputs even when working with very short trajectories and in the presence of experimental noise. We further demonstrate the application of transfer learning to experimental and simulated data not included in the training/test dataset. This allows for a full, high-accuracy characterization of experimental trajectories without the need of any prior information. 
\end{abstract}

\maketitle

\begin{figure*}
	\begin{center}		
		\includegraphics[width=2\columnwidth]{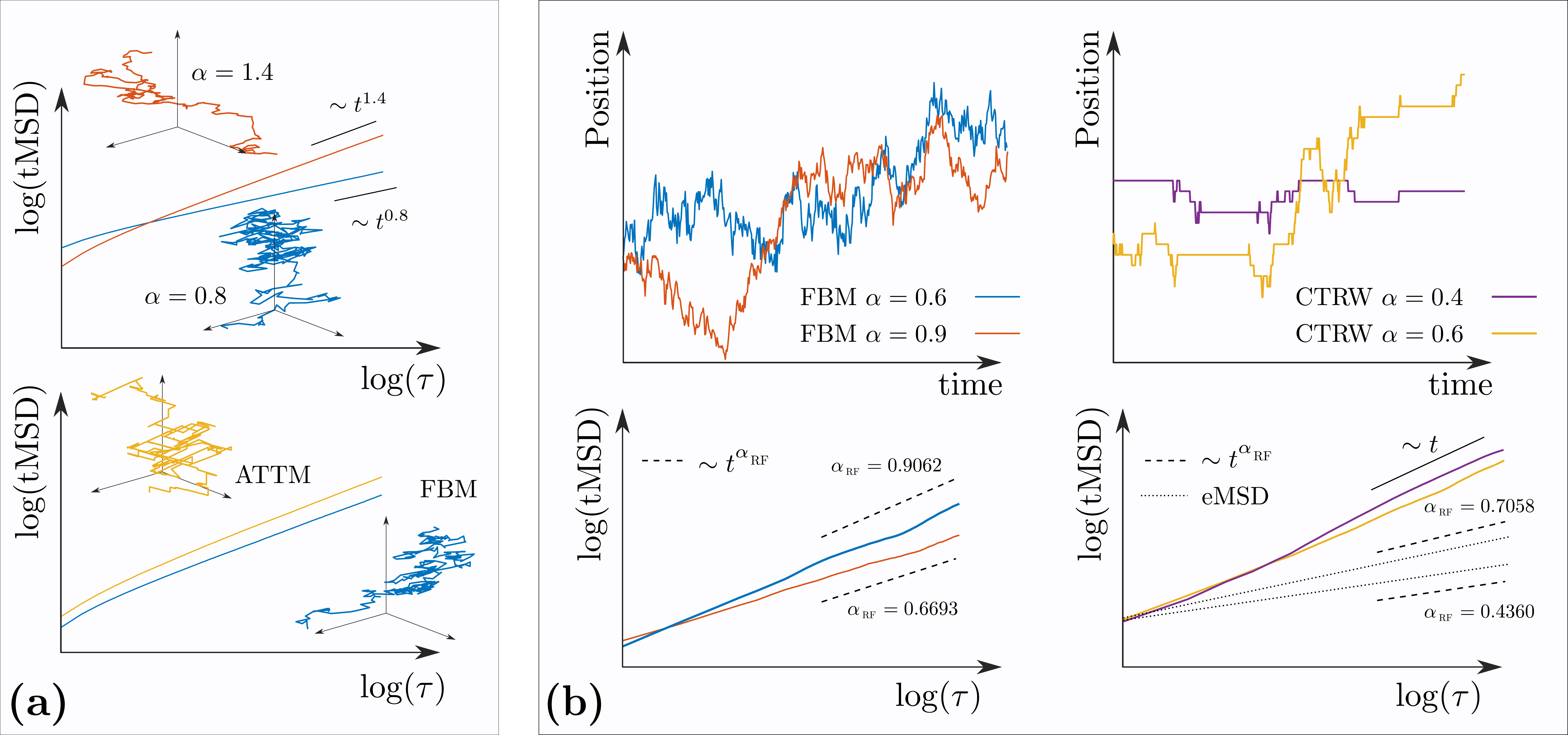}
		
		\caption{\textbf{Examples of single molecule trajectory heterogeneity} 
		\textbf{(a)} Time-averaged mean squared displacement (tMSD) calculated from single trajectories. In the upper panel, we show the tMSD of two trajectories corresponding to molecules that display different anomalous exponents $\alpha$ in spite of belonging to the same physical process.  In the lower panel,  we show the tMSD of two trajectories generated from different diffusion processes but producing a similar $\alpha$. In this case, the exponent determined from the tMSD cannot be used to discriminate among models.
		\textbf{(b)} The upper panels display representative trajectories generated from FBM (left) and CTRW (right) models. The two bottom panels show the corresponding tMSD curves (continuous lines) using the same color coding. Since the subdiffusive FBM is ergodic, the tMSD of a single trajectory can be used to extract the anomalous exponent of the model. However, for nonergodic processes such as the CTRW, the tMSD of a single trajectory is a random variable and its anomalous exponent can be different from the one associated to the model, which calculation requires the use of ensemble averaging. In both cases, the RF algorithm is able to determine the correct anomalous exponent associated with the model, even when it does not simply correspond to the exponent of the single trajectory tMSD, as in the nonergodic case. The values of exponent provided by the RF algorithm are schematically represented by the dashed lines, and are in good agreement with the ground truth values, which for the CTRW were also calculated by the eMSD (dotted lines). \label{fig:fignew}}		
	\end{center}
	
\end{figure*}

In the last decades, the research in biophysics has conveyed large efforts toward the development of experimental techniques allowing the visualization of biological processes one molecule at a time~\cite{Miller2018,Sahl2017,Manzo2015review, Hofling2013}. These efforts have been mainly driven by the concept that ensemble-averaging hides important features that are relevant for cellular function. Somehow expectedly, experiments performed by means of these techniques have shown a large heterogeneity in the behavior of biological molecules, thus fully justifying the use of these raffinate tools. Besides, experiments performed using single particle tracking~\cite{Manzo2015review} have revealed that even chemically-identical molecules in biological media can display very different behaviors, as a consequence of the complex environment where diffusion takes place.  By way of example, this heterogeneity is reflected in the broad distribution of dynamic parameters of distinct individual trajectories corresponding to the same molecular species, such as the diffusion coefficient, well above stochastic indetermination. Typically, the trajectories are analyzed by quantifying the (time-averaged) mean square displacement (tMSD) as a function of the time lag $\tau$~\cite{Metzler2014}:
\begin{equation}
\overline{\delta^2(\tau)}=\frac{1}{t_{max}-\tau}\int_0^{t_{max}-\tau} [ x(t' +\tau) -x(t')]^2  dt'.
\end{equation}
The calculation of this quantity - expected to scale linearly for a Brownian walker in a homogeneous environment -  has provided a ubiquitous evidence of anomalous behaviors in biological systems, characterized by an asymptotic nonlinear scaling of the tMSD curve $\overline{\delta^2} \sim \tau^\alpha$. More experiments have shown that the anomalous exponent can vary from particle to particle (Fig.~\ref{fig:fignew}(a)) as a consequence of molecular interactions and that these changes can be experienced by the same particle in space/time~\cite{Weron2017}. Several methods have been proposed to accurately estimate this exponent for single trajectories~\cite{Kepten2013, Burnecki2015} in the presence of experimental limitations, such as optical diffraction and the finite length of the trajectory.

In some cases, this heterogeneity leads to exotic effects, such as the breaking of ergodicity observed in several cellular systems~\cite{Jeon2011, Weigel2011,Tabei2013, Manzo2015}. Nonergodicity implies the nonequivalence of time and ensemble averages of the mean squared displacement (MSD). In the nonergodic case, $\overline{\delta^2(\tau)}$  remains random even in the long measurement times, i.e., the diffusion coefficients are irreproducible but the distribution of the tMSD is universal~\cite{Akimoto2016}. For nonergodic models, the determination of the anomalous exponent at the single trajectory level does not fully characterize the model, since the tMSD can have different scaling when averaged with respect to the time or to the ensemble (Fig.~\ref{fig:fignew}(b)). Therefore, it requires the calculation of the ensemble-averaged MSD (eMSD) over a rather large number of trajectories~\cite{Metzler2014}.

The emergence of anomalous behavior has also been widely studied from the theoretical point of view and conceptually-different models have been proposed~\cite{Metzler2014}. However, the fact that models with different physical properties can produce the same tMSD exponent, strongly limits the unambiguous determination of the underlying dynamics, based only on the evaluation of the tMSD (Fig.~\ref{fig:fignew}(a)). In order to solve this ambiguity, a large effort has been made to classify experimental data showing anomalous transport. As an example, the use of alternative estimators~\cite{Magdziarz2009, Magdziarz2011} has been proposed to determine whether the pioneering results of Golding and Cox~\cite{Golding2006} were arising from a continuous-time random walk (CTRW)~\cite{Sokolov2011} or fractional Brownian motion (FBM)~\cite{Mandelbrot1968}. This search for a better classification between CTRW and FBM often relied in the determination of the (non)ergodicity of the data~\cite{Deng2009, Magdziarz2011, Meroz2015, Lanoiselee2016}, since CTRW is consistent with weak ergodicity breaking~\cite{Bouchaud1992}. The experimental evidence of nonergodic behavior has boosted the proposal of new theoretical frameworks consistent with these features~\cite{Massignan2014, Chubynsky2014}. 

In this scenario, determining whether a single-molecule trajectory (or a short segment of it) displays normal or anomalous behavior by its tMSD scaling exponent and associating the motion to a specific physical model are elements of paramount importance to gain insight about the biophysical mechanism underlying anomalous diffusion, thus providing a detailed picture of a variety of phenomena. Recent works in this direction have focused on classification schemes based on optimization procedures~\cite{Burnecki2015}, Power Spectral Density~\cite{Krapf2018}, or Bayesian approaches~\cite{Hinsen2016, Thapa2018}.

\begin{figure*}
	\begin{center}		
		\includegraphics[width=2\columnwidth]{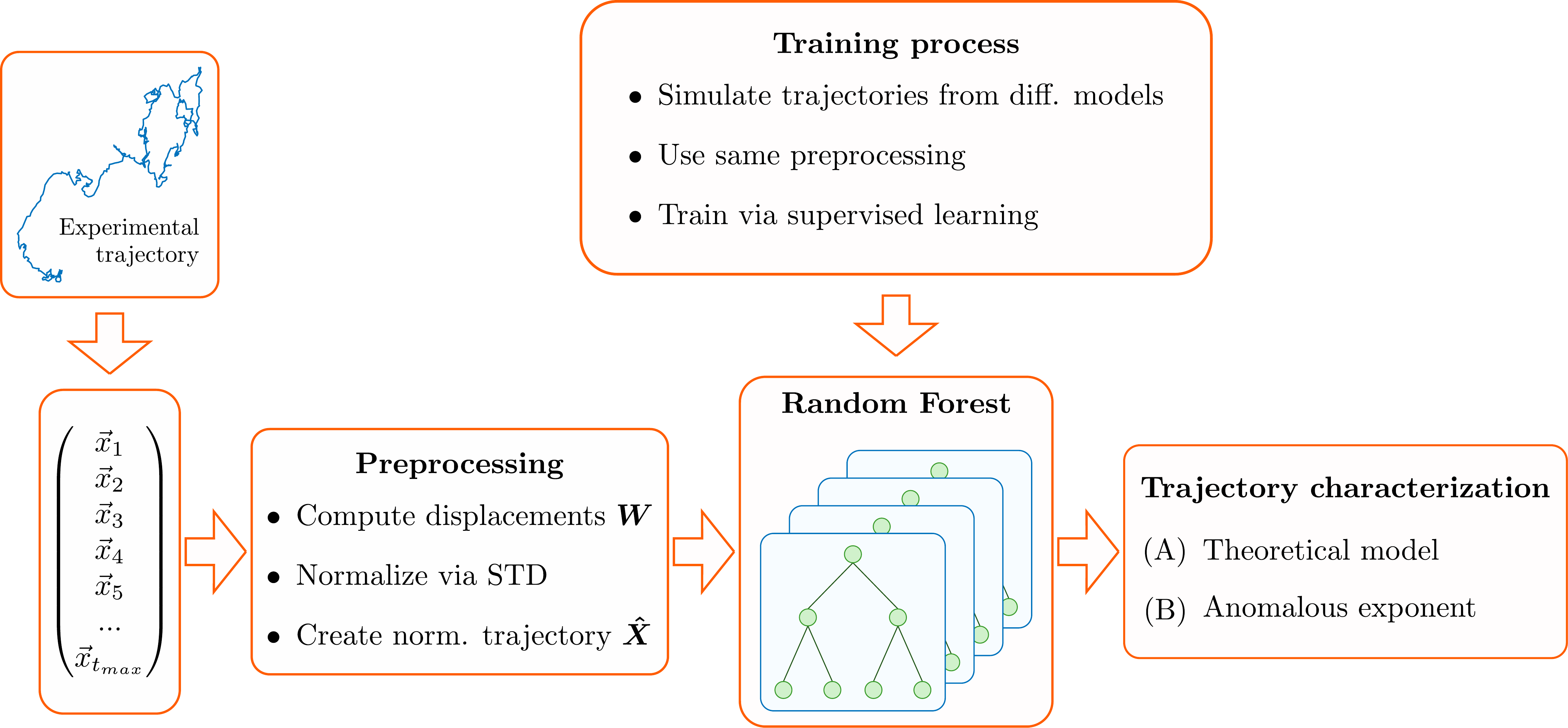}
		\caption{\textbf{Schematic of the method.} An experimental trajectory is first transformed into a time series and preprocesses according to the procedure described in Section~\ref{sec:model_dataset_preprocessing}. The trajectory is fed to the algorithm for its characterization through a RF, previously trained on simulated data. We show RF capability to extract two characteristics of the trajectory: (A) the most likely theoretical model among the ones contained in the training dataset and (B) the anomalous exponent $\alpha$. \label{fig:fig0}}		
	\end{center}
	
\end{figure*}

Surprisingly, in spite of the fast rise of Machine Learning methods, little efforts have been made to characterize single trajectories. A few attempts in this sense have been mainly directed to qualitatively discriminate among confined, anomalous, normal or directed motion~\cite{Dosset2016, Kowalek2019}, without extracting quantitative information of classifying with respect to the underlying physical model. 

This paper presents a Machine Learning algorithm based on the Random Forest (RF) architecture that efficiently and robustly characterizes single trajectories at different levels: first, obtaining the discrimination among several diffusion models; then, providing the estimation of the exponent that characterizes the anomalous diffusion, thus inherently classifying between normal and anomalous (sub- and super-) diffusion. The algorithm allows to accurately tackle these challenging problems even when dealing with short and noisy trajectories. 

In comparison to previously reported methods, the algorithm we present in this paper allows for the characterization of a trajectory without prior knowledge about the process from which the single trajectory has been extracted.  Most of the works existent in the literature focus their studies on anomalous ergodic trajectories (usually FBM related processes), whereas our method is robust against the appearance of nonergodicity. While showing similar accuracy as others on ergodic trajectories~\cite{Thapa2018, Burnecki2015}, to the best of our knowledge, our method represents the first attempt to extract the anomalous exponent for non-ergodic processes through single-trajectory characterization.


\section{Machine learning method}
\label{sec:model_dataset_preprocessing}

In this section, we outline the main parts of the trajectory characterization algorithm, consisting of: the Machine Learning algorithm that takes the form of a RF architecture; the simulated  dataset; and the preprocessing applied to the dataset before being analyzed by the RF. Figure~\ref{fig:fig0} shows a schematic representation of the pipeline.

\paragraph*{Random Forest}
RF is an architecture based on Decision Trees. A decision tree is an efficient non-parametric method widely used for classification and regression problems~\cite{Breiman1984}. The basic idea consists in producing recursive binary splits of the input space, so that the samples with the same label are grouped together. The criterion to produce the splits is based on a homogeneity measure (usually, the information entropy) of the target variable within each of the obtained groups. 
In regression problems, a commonly used criterion is to select the split that minimizes the Mean Squared Error (MSE); this recursive process continues until some stopping rule is satisfied, e.g., a common one is to consider that a tree node can be split if it contains more than a given number of samples; therefore, the minimum number of samples required to split a tree node should be adjusted in order to control the size of the tree, thus preventing overfitting.
Once a decision tree is obtained, the output for unseen samples is computed just passing them through the nodes of the tree, where a decision is made with respect to which direction to take. Finally, a terminal tree node is reached, where the output is obtained.

A RF is a tree-based ensemble method, which builds several decision tree models independently and then computes a final prediction by combining the outputs of the different individual trees~\cite{Breiman2001}. In particular, the ensemble is produced with single trees built from samples drawn randomly with replacement (bootstrap) from the training set. An additional randomness is added when splitting a tree node because the split is chosen among a random subset of the input variables, selected in this case without replacement, instead of the greedy approach of considering all the input variables. Due to this randomization, the bias of the ensemble is slightly higher than that of a single tree, but the variance is decreased and the model is more robust to variations in the dataset.

RF is a very powerful, state-of-the-art technique for both regression and classification problems, usually outperforming not only single decision trees but also sophisticated models, as shown in a thorough comparison study~\cite{JMLR:v15:delgado14a}.

\paragraph*{Training and test datasets}
The training dataset is built out of numerical simulations of trajectories from various kinds of theoretical models. As a natural choice, we included three of the best-known and used models that can give rise to anomalous diffusion: CTRW~\cite{Sokolov2011}, FBM~\cite{Mandelbrot1968} and L\'evy walks (LW)~\cite{Zaburdaev2015}. In addition, we included trajectories from the annealed transient time model (ATTM)~\cite{Massignan2014}.  In the ATTM, a diffuser performs a random walk but stochastically changes the diffusion coefficient at random times. Both the diffusion coefficient and the time at which the diffusivity changes are drawn from distributions with a power law behavior~\cite{Massignan2014}. Its time-averaged MSD shows a linear scaling, but the model has a regime in which it displays weak ergodicity breaking. The ATTM has been shown to reproduce the features observed for the diffusion of a cell membrane receptor~\cite{Manzo2015},  one of the experimental datasets analyzed.

\paragraph*{Preprocessing}
Our aim is to design a method that can be used to accurately characterize heterogeneous trajectories without having to calculate other parameters or using a priori knowledge. In order to be able to analyze data coming from any possible spatiotemporal scale, we designed a preprocessing procedure that properly rescale the data. We implemented the following procedure, chosen among other plausible normalization techniques as it gives rise to the best results in terms of accuracy: 
\begin{enumerate}
	\item We use one of the models above to simulate the trajectory of a particle during $t_{\max}$ time steps. The result is a vector of positions, $\bm{X} = (\bm{x}_1,\bm{x}_2,...,\bm{x}_{t_{\max}})$.
	\item This vector is transformed into a vector of distances traveled in an interval of time $T_{\rm{lag}}$, i.e.,   $\bm{W} = (\Delta\bm{ x}_1,\Delta\bm{ x}_2,...,\Delta \bm{x}_{J-1})$,, where $J=t_{\max}/T_{\rm{lag}}$. We define $\Delta\bm{ x}_i$ as
	\begin{equation}
	\Delta\bm{ x}_i = \left| \bm{x}_{iT_{\rm{lag}}}-\bm{x}_{(i+1)T_{\rm{lag}}} \right|.
	\end{equation}
	\item To normalize the data, we divide $\bm{W}$ by its standard deviation (STD) to get a new vector $\hat{\bm{W}}$ .
	\item Then, we do a cumulative sum of $\hat{\bm{W}}$ to construct a normalized trajectory $\hat{\bm{X}}$. 
\end{enumerate}

Summarizing, the previous procedure generates a new trajectory which is constructed via the normalized displacements of the original trajectory. This makes that the magnitudes of the resulting trajectories are comparable, no matter what were their original values. While the RF could be trained using $\hat{\bm{W}}$, our results show that training with $\hat{\bm{X}}$ gives indeed much better results. The same preprocessing is applied to both the simulated and experimental trajectories used in Sections~\ref{sec:benchmarking} and~\ref{sec:transfer}.

\begin{figure*}
	\includegraphics[width=2\columnwidth]{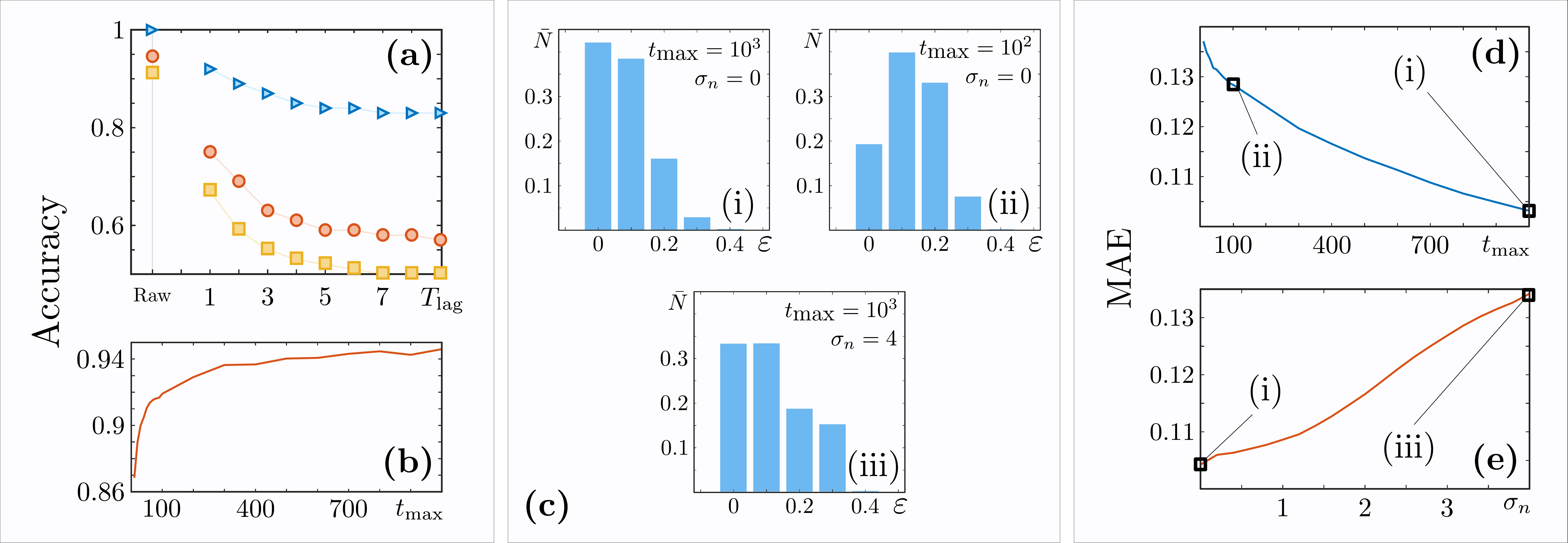}
	\caption{\textbf{Benchmarking the RF algorithm:} \textbf{(a)} Accuracy of the RF when discriminating among models as a function of the preprocessing parameter $T_{\mbox{lag}}$. Blue triangles = CTRW {\it vs} FBM, Red circles= CTRW, LW, FBM, ATTM, and Yellow squares= CTRW {\it vs} ATTM. 
	\textbf{(b)} Accuracy of the model discrimination as a function of the length of the trajectories $t_{max}$.
	\textbf{(c)} Histograms of the error in the prediction of the anomalous exponent for different values of the length trajectory $t_{\mbox{max}}$ and noise variance $\sigma_n$. Y-axis is percentage of trajectories $\bar{N}$ with given error $\varepsilon$ when predicting the value of $\alpha$. Each histogram corresponds to the highlighted points in (d) and (e).
	\textbf{(d)} MAE of the anomalous exponent prediction as a function of the length of the trajectories $t_{max}$.
	\textbf{(e)} MAE in anomalous exponent prediction as a function of the variance of the normal noise variance. The RF was trained with a dataset without noise and then asked to classify trajectory of such models with normal noise given by~\eqref{eq:noise}. $T_{\rm{lag}}=0$, i.e. raw trajectories, were used for (b), (c), (d) and (e).}
	\label{fig:benchmarking}
\end{figure*}	


\section{Trajectory characterization as a machine learning problem}
\label{sec:benchmarking}
We will use our method to characterize single trajectories according to two different schemes: (A) discrimination among diffusion models; (B) prediction of the anomalous exponent $\alpha$, that inherently implies classification as normal or anomalous diffusion. For each of these problems, we created a dataset of $1,2\cdot10^5$ trajectories with $t_{max}=10^3$, divided into a training and test set with ratio 0.8/0.2, respectively. The results presented in all the figures and the values of the accuracy discussed in the text correspond to the ones measured in the test set, ensuring that the RF does not present overfitting in any of the problems considered. The different classes considered in each problem have an equal number of trajectories, hence allowing us to use the accuracy as a measure of the goodness of the RF. For technical details and an practical example of the implementation, we refer the reader to the Supplementary Material and to the code repository Ref.~\cite{github}. 

\paragraph*{A. Discrimination among diffusion models}
In order to predict the diffusion model underlying a certain trajectory, we construct a RF whose input is the normalized trajectory $\hat{\bm{X}}$, and the output is a number between 0 and $N-1$ corresponding to the different models,  with $N$ the total number of models used in the training. Figure~\ref{fig:benchmarking}(a) shows the accuracy of the RF.  Each line corresponds to a training dataset made up of different models.  In the absence of data preprocessing (point marked as 'Raw' in the x-axis), the RF shows large accuracy. The accuracy drops significantly as $T_{\rm{lag}}$ increases, likely as a consequence of the removal of microscopical properties of the model, such as short-time correlations, hence preventing the RF from learning important features of them. This might lead to the conclusion that the filtering introduced by the preprocessing steps only limits the time resolution.  This is obviously true for simulated data, obtained at the same scale, for which preprocessing is unnecessary. However, when dealing with experimental data of unknown spatiotemporal scale, such a preprocessing is of fundamental importance to be able to apply the same architecture and training dataset, in spite of the little loss of performance.

In addition, the accuracy heavily depends on similarities among the models to be discriminated. For example, the accuracy obtained  with a RF trained only with trajectories reproducing conceptually different models such as FBM and CTRW  (triangular markers in Fig.~\ref{fig:benchmarking}(a)) is higher than the one obtained when including in the training models with similar characteristics, such as CTRW and ATTM, independently of $T_{\rm{lag}}$ (red circles and yellow squares in Fig.~\ref{fig:benchmarking}(a)).

\paragraph*{B. Anomalous exponent estimation}
A first approximation toward the characterization of the anomalous exponent can be based on a regression problem, in which the output of the RF is the value of the anomalous exponent $\alpha$. The nature of the regression algorithm makes that the output of the RF is the continuous value which better satisfies the constraints learn during training.

To characterize the performance of the method, we calculate the prediction error $\varepsilon$ of a trajectory as the absolute distance between the predicted exponent and the ground truth value. The percentage of trajectories $\bar{N}(\varepsilon)$ with a given error $\varepsilon$ is represented in the bar plots of Fig.~\ref{fig:benchmarking}(c) for three different cases and a subdiffusive dataset including trajectories obtained from FBM, CTRW and ATTM. The case (i) considers trajectories with $t_{max}=10^3$ without noise, while the case (ii) and (iii) show results for shorter and noisy trajectories (see discussion below). For case (i) the calculated  mean absolute error (MAE) of the prediction of the anomalous exponent gives a value of $0.11$. Moreover, the histogram showed in Fig.~\ref{fig:benchmarking}(c)(i) shows that for $\sim 80\%$ of the trajectories, the output exponent lies within $0.1$ from the true value.

\paragraph*{C. Experimental scenario: short and noisy trajectories}

A remarkable feature of the method is the possibility to correctly characterize very short trajectories. In Fig~\ref{fig:benchmarking}(b) and (d), we show the ability of the RF to characterize short trajectories. In Fig~\ref{fig:benchmarking}(b), we plot the accuracy in model discrimination as a function of the length of the trajectories, $t_{max}$. In Fig~\ref{fig:benchmarking}(d), a similar study is done, now tracking the MAE of the RF trained to predict $\alpha$. Although we observe an expected decrease of performance for short trajectories, both plots show that the RF is able to characterize trajectories as short as only 10 points. Quantitatively, when comparing trajectories of 10 points with larger ones, of 1000 points, the model discrimination accuracy only decreases by a factor of $8.2$\%, while the MAE decreases by a factor of 18\%. Panel (ii) in Fig~\ref{fig:benchmarking}(e) shows the error distribution when predicting $\alpha$ for $t_{max}=100$.

Importantly, one has also to take into account that the experimental trajectories have a limited localization precision,  that results into Gaussian noise. Therefore, it is necessary to test the robustness to noise of the RF. To this end, we trained the RF with trajectories simulated as described before and then we try to predict the anomalous exponent of trajectories belonging to the same dataset, but whose positions $\bm{X}$ were perturbed with noise to obtain the dataset $\bm{X}^{(n)}$
\begin{equation}
\bm{x}^{(n)}_i = \bm{x}_i + \mu_i(\mu, \sigma_n),
\label{eq:noise}
\end{equation}
where $\mu_i(\mu, \sigma_n)$ is a random number retrieved from a Normal distribution with mean $\mu = 0$ and variance $\sigma_n$. The results obtained for training with FBM, CTRW and ATTMs are presented in Fig.~\ref{fig:benchmarking}(d). The RF shows a great robustness against noise. For $\sigma_n<1$, the MAE appears almost unaffected. When increasing $\sigma_n$, we see that the MAE increases, as expected, but even for large $\sigma_n$ the MAE is still reasonable.


\section{Transfer learning in simulated and experimental data}
\label{sec:transfer}

To further show the advantages of our Machine Learning algorithm, we applied it to three sets of trajectories different from those included in the training/test dataset.  This is often referred as transfer learning, as certain architecture is trained in one setting and then applied to a different one. For this, we will consider three datasets:

\begin{enumerate}[label=(\roman*)]
	\item Simulated data coming from a recently presented model~\cite{Munoz2018},  describing the movement of a diffuser in a network of compartments of random size and random permeability, both drawn from universal distributions. This model shares the same subordination as the quenched trap model, i.e. a CTRW with power-law distributed trapping times and recapitulates the complexity and heterogeneity found in some biological environments. This choice allows to test the algorithm over a conceptually different model with respect to the training dataset, while having the advantage of tuning  the value of  anomalous exponents.
	\item Experiment 1, reporting the motion of individual  mRNA molecules inside live bacterial cells~\cite{Golding2006}. The tMSD shows anomalous diffusion with $\alpha\sim0.7$; this behavior has been associated to FBM~\cite{Magdziarz2009, MolinaGarcia2016}.
	\item Experiment 2, corresponding to a set of trajectories obtained for the diffusion of a membrane receptor in living cells~\cite{Manzo2015}.  Although the time-averaged MSD shows a nearly linear behavior, the data present features of ergodicity breaking due to changes of diffusivity~\cite{Balcerek2019} and have been associated to the ATTM model.
	
\end{enumerate}

Following the scheme presented in Fig.~\ref{fig:fig0}, first we train the RF with simulated trajectories obtained with different theoretical frameworks. It should be noted that for this section, since we deal with trajectories that do not show superdiffusive behavior, we do not include the L\'evy walks process in the training dataset.

\begin{figure}
	\includegraphics[width=1\columnwidth]{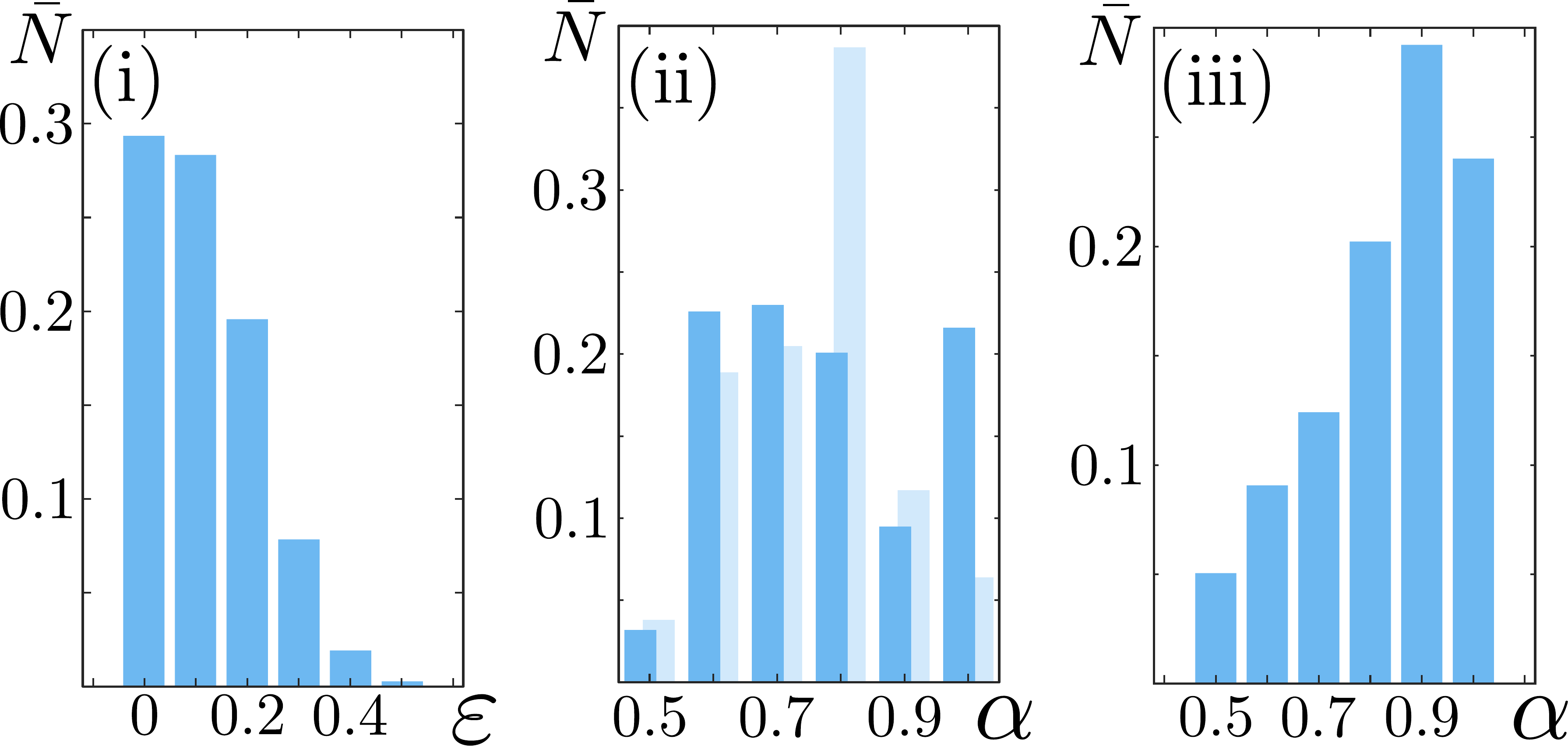}
	\begin{center}
		\caption{\textbf{Transfer learning: predicting the anomalous exponent for experimental trajectories.} Labels (i),(ii) and (iii) refer to the datasets discussed in Section~\ref{sec:transfer}. For dataset (i), we plot the percentage of trajectories $\bar{N}$ where the predicted value of $\alpha$ has an absolute error $\varepsilon$. As it is a simulated dataset, exponents from 0.2 to 1 are considered. The input to the RF were the raw trajectories, with no preprocessing. For datasets (ii) and (iii), we present the percentage of trajectories predicted to have an anomalous exponent $\alpha$. The trajectories were preprocessed with $T_{\rm{lag}}=1$. For dataset (ii), we present results two training datasets: dark blue for a mixed dataset and light blue for a FBM dataset. }
		\label{fig:experiments}
	\end{center} 	
\end{figure}

\paragraph*{Results}

Following the same structure of the previous section, we start by discriminating the diffusion model that can be associated to datasets (i)-(iii).  The results are reported in Table~\ref{table}, showing a high rate of correct classification for the dataset (i). For the experimental data in datasets (ii) and (iii), we do not dispose of ground truth values, thus we compare our results with those of previous analysis, performed with alternative methods. For the trajectories of Experiment 1, we found that the algorithm largely assign them to the FBM, in strong agreement with previously reported results based on the concept of variation~\cite{Magdziarz2009}. The data of Experiment 2 are mainly assigned to the ATTM model. This model was shown to reproduce features observed in these data, such as subdiffusion and weak ergodicity breaking~\cite{Manzo2015}. Moreover, a little fraction of trajectories are classified as CTRW. As previously mentioned, CTRW and ATTM share similar features (such as time subordination), increasing the difficulty in discriminating between them. This appears to be the main source of error in the results. 

\begin{table}[h]
	
	\centering
	\caption{Process discrimination for the datasets considered in section \ref{sec:transfer}. Shown is the percentage of trajectories classified as associated to each model. The results for (i) were done with $T_{\rm{lag}}=0$ and for datasets (ii) and (iii) with $T_{\rm{lag}}=1$.}
	\begin{tabular}{llll}
		\hline
		\multicolumn{1}{c}{Dataset} & \multicolumn{3}{c}{Predicted Model}                 \\ \hline
		& CTRW            & FBM             & ATTM            \\ \hline
		(i) Compartments model                        & \textbf{89.2\%} & 0               & 10.7\%          \\
		(ii)  Experiment 1                     & 4.5\%           & \textbf{86.6\%} & 8.9\%           \\ 
		(iii)  Experiment 2                      & 16.4\%          & 33.2\%          & \textbf{50.4\%} \\
		\hline
	\end{tabular}
	
	\label{table}
\end{table}

To obtain further insights on the study of the diffusion, we used the RF to extract the anomalous exponents. For the first dataset (i), based on simulations, we generated trajectories having a broad range of subdiffusive trajectories, namely $\alpha\in[0.2,1]$. Then, we used the trained RF to predict the value of the anomalous exponent and evaluated the error as the absolute value of the difference between the actual and predicted $\alpha$.  The results are reported in the histogram of Fig.~\ref{fig:experiments} (i) and display a distribution similar to the one obtained for the training/testing data.  Thus, we run the same procedure on the experimental data. For the two datasets, in Fig.~\ref{fig:experiments}(ii)-(iii) we report the values obtained for the anomalous exponent $\alpha$. The histogram of the $\alpha$ obtained for the trajectories of Experiment 1 (dark blue) shows mainly subdiffusive values, peaked in the range $0.6-0.8$. This is in good agreement with the original paper~\cite{Golding2006}, where $\alpha$ was estimated by means of two different approaches as $0.7$ and $0.77$. However, the method also classifies a percentage of the trajectories as having $\alpha=1$. 
Importantly, the performance of the method can be further improved by taking advantage of the results of the model discrimination discussed above and shown in Table~\ref{table}. In fact, when the latter classification indicates that most of the trajectories follow a specific diffusive model, one can train the algorithm with a dataset composed only of trajectories simulated with that model. This kind of training produces exponent values in the same range, but largely reduce the fraction of those associated to $\alpha=1$, as shown in Fig.~\ref{fig:experiments}(ii) (light blue).

Last,  in Fig.~\ref{fig:experiments}(iii) we plot the distribution of exponents obtained for the Experiment 2. The subdiffusive values show a large number of occurrences in the $0.8-0.9$ range, compatible with the exponent $0.84$ calculated in previous studies~\cite{Manzo2015}.  Noteworthy, due to the nonergodic nature of the data, in the original paper $\alpha$ could only be calculated from the ensemble-averaged MSD, whereas the RF is able to determine this exponent from single trajectories.

\section{Conclusions}
\label{sec:conclusions}
We have presented a machine learning method, based on a Random Forest architecture, which is capable to analyze a single trajectory and to determine the theoretical model that describes it at best. Moreover the same method is used for predicting its anomalous exponent with high accuracy, and thus classify the motion as normal or anomalous. The method does not need any prior information over the nature of the system from which the trajectory is obtained. It acts as a blackbox, which we train with a dataset of simulated trajectories, and then it is used to characterize the trajectory of interest. In particular, its spatial scale is not of any relevance, as  we devised a preprocessing strategy which rescales trajectories to obtain comparable estimators from very different systems. The method requires a minimal amount of information. First, because it performs extremely well even with surprisingly short trajectories. Second, because it is robust with respect to the presence of a large amount of thermal noise, and can thus be applied even with low localization precision. We showcase the suitability of our method by applying it to two experimental datasets by means of transfer learning. Overall, this method can be of large application to characterize experiments from several research areas. In contrast to other methods, it can determine the type of diffusion and the anomalous exponent also for nonergodic models, without the need of performing ensemble averages. 
We note that recent works show how other Machine Learning architectures, such as convolutional neural networks and long-short term memory neural networks, are also capable of doing single trajectory characterization~\cite{Kowalek2019, Bo2019}. The development of these methods and of other deep learning architectures may help to avoid the preprocessing procedure and could lead to increase the accuracy on the problems described in this work.

\section*{Acknowledgments}
This work has been funded by the Spanish Ministry MINECO (National Plan15 Grant: FISICATEAMO No. FIS2016-79508-P, SEVERO OCHOA No. SEV-2015-0522, FPI), European Social Fund, Fundaci\'o  Cellex, Generalitat de Catalunya (AGAUR Grant No. 2017 SGR 1341 and CERCA/Program), ERC AdG OSYRIS, EU FETPRO QUIC, and the National Science Centre, Poland-Symfonia Grant No. 2016/20/W/ST4/00314. C.M. acknowledges funding from the Spanish Ministry of Economy and Competitiveness and the European Social Fund through the Ram\'{o}n y Cajal program 2015 (RYC-2015-17896) and the BFU2017-85693-R and from the Generalitat de Catalunya (AGAUR Grant No. 2017SGR940). G.M. acknowledges financial support from Fundació Social La Caixa. MAGM acknowledges funding from the Spanish Ministry of Education and Vocational Training (MEFP) through the Beatriz Galindo program 2018 (BEAGAL18/00203). We gratefully acknowledge the support of NVIDIA Corporation with the donation of the Titan Xp GPU.

\section*{Supplementary materials}
 \paragraph*{Hyperparameters.}
 We summarize here the different parameters used to train the RF models whose results are presented in this work. Common to all are the number of ensembled trees (100), the train/test dataset ratio (0.8/0.2) and the size of training set ($1,2\cdot 10^5$ trajectories). For the details of each figure, see Table~\ref{table:hyperparameters}.

 \begin{table*}[]
 	\begin{tabular}{lccc}
 		& Models in dataset & Range of $\alpha$'s                   & $t_{max}$  \\ \hline
 		Figure \ref{fig:benchmarking}(a)                & See caption  & {[}0.2, 2{]}   & $10^3$     \\
 		Figure \ref{fig:benchmarking}(b) & CTRW/LW/FBM/ATTM  & {[}0.2, 2{]}   &  See x-axis     \\
 		Figure \ref{fig:benchmarking}(c)(i)                & CTRW/LW/FBM/ATTM  & {[}0.5, 1{]}   & $10^3$     \\
 		Figure \ref{fig:benchmarking}(c)(ii)                & CTRW/LW/FBM/ATTM  & {[}0.5, 1{]}   & $10^2$     \\
 		Figure \ref{fig:benchmarking}(c)(iii)                & CTRW/LW/FBM/ATTM  & {[}0.5, 1{]}   & $10^3$     \\
 		Figure \ref{fig:benchmarking}(d)                & CTRW/LW/FBM/ATTM  & {[}0.5, 1{]}   & See x-axis \\
 		Figure \ref{fig:benchmarking}(e)                & CTRW/LW/FBM/ATTM        & {[}0.5, 1{]}   & $10^3$     \\
 		Figure \ref{fig:experiments}(i)             & CTRW/FBM/ATTM     & {[}0.5, 1{]}   & $10^3$     \\
 		Figure \ref{fig:experiments}(ii) dark blue  & CTRW/FBM/ATTM     & {[}0.5, 1.2{]}  & $10^3$       \\
 		Figure \ref{fig:experiments}(ii) light blue & FBM               & {[}0.5, 1.2{]}  & 300        \\
 		Figure \ref{fig:experiments}(ii) dark blue  & CTRW/FBM/ATTM     & {[}0.5, 1.2{]} & 200  
 		
 	\end{tabular}
 	
 	\caption{Hyperparameters for the training of the different models exposed in this work.}
 	\label{table:hyperparameters} 
 \end{table*}

\end{document}